

\magnification=1200
\baselineskip=20pt
\centerline{\bf On Diff($S^1$) Covariantization Of}
\centerline{\bf Pseudodifferential Operator}
\vskip 1.5cm
\centerline{Wen-Jui Huang}
\vskip 1.0cm
\centerline{Department of Physics}
\centerline{National Tsing Hua University}
\centerline{Hsinchu, Taiwan R.O.C.}
\vskip 1.5cm
\centerline{\bf \underbar{Abstract}}
\vskip 1cm

A study of diff($S^1$) covariant properties of pseudodifferential operator
of integer degree is presented. First, it is shown that the action of
diff($S^1$) defines a hamiltonian flow defined by the second
Gelfand-Dickey bracket if
and only if the pseudodifferential operator transforms covariantly. Secondly,
the covariant form of a pseudodifferential operator of degree $n \neq 0,
\pm 1$	is constructed by exploiting the inverse of covariant derivative. This,
in
particular, implies the existence of primary basis for
$W_{KP}^{(n)}$ ($n \neq
0, \pm 1)$.

\vfil\eject

\noindent{\bf \underbar{I. Introduction}}
\vskip .5cm

It has been known that the Lax formulation [1-3]
of integrable systems provides a
very useful method to construct W-type algebras. For example, the so called
$W_n$ algebra[4] is nothing but the
second Gelfand-Dickey bracket associated with
the differential operator[5]
$$\Lambda_n = \partial^n + u_2 \partial^{n-1} + \dots +u_n \eqno(1.1)$$
Recently, the study of the hamiltonian structures of the KP hierarchy[6-9]
leads to
the consideration of the Gelfand-Dickey brackets associated with the
pseudodifferential operators of the form[10-13]
$$L_n = \partial^n + u_2 \partial^{n-1} + \dots + u_p \partial^{n-p} + \dots
\eqno(1.2)$$
It is now known that the second Gelfand-Dickey bracket indeed defines a
hamiltonian structure provided $n \neq 0$. Interestingly, it is possible to
generalize further to the case when $n$ is a complex number $q$.
This generalization
  leads to a large class of W-type algebras called $W_{KP}^{(q)}$[14].

The importance of these W-type algebras comes from the fact that they all
contain a Virasoro subalgebra and a set of generators of spin higher than 2.
For instance, for the $W_n$ algebra, which is associated with the differential
operator (1.1), $u_2$ is the Virasoro generator and there is a differential
algebra automorphism $\{ u_j \}_{j \geq 3} \rightarrow \{ w_j \}_{j \geq 3}$
such that $w_j$ is a primary generator of spin $j$[15-17]. The proof of
the last
statement relies on the equivalence of the diff($S^1$)-covariance and the
Virasoro hamiltonian flow defined by the second Gelfand-Dickey bracket and
on the possibility of covariantizing the differential operator. The set of
generators $\{ u_2, w_3, \dots, w_n \}$ is called
the primary basis for $W_n$.
It is believed that, for $n \neq 0, \pm 1$,
a primary basis for $W_{KP}^{(n)}$
should also exist[14,18].
However, to the best of our knowledge, an explicit proof is
still lacking. It is the purpose of this paper to explicitly covariantize the
pseudodifferential operators and thus provides a proof of the existence of
primary basis.

We organize this paper as follows. In Sec. II, we collect basic definitions and
concepts. In Sec. III,	we prove the
equivalence of the diff($S^1$)-covariance
and the Virasoro hamiltonian flow. In Sec. IV,	the covariantization of the
pseudodifferential operators is carried out. In Sec. V, the algebra
$W_{KP}^{(n)}$ with $n \geq 2$ is studied and, in particular,
the first few primary generators of spin higher than $n$ are explicitly
defined. Finally, we present discussions and conclusions in Sec. VI.
\medskip

\noindent{\bf II. \underbar{Basic Definitions}}

We consider the differential operator
$\partial = {\partial \over {\partial x}}$.
The powers of this operator satisfy
$$\eqalign{ \partial^0 &= 1 \cr
 \partial^i \partial^j &= \partial^{i+j} \cr
 \partial^i f &= \sum^{\infty}_{ k=0} \pmatrix{i \cr k}  f^{(k)}
\partial^{i-k}\cr}
 \eqno(2.1)$$
where $f^{(k)}$ is the k-th derivative of the function $f(x)$ and
$$\eqalign{  \pmatrix{ i \cr k} &= {{i(i-1)
\dots (i-k+1)} \over k!} \cr
\pmatrix{i \cr 0}  &= 1 \cr}\eqno(2.2)$$
Quite often we include a subscript (e.g. $\partial_t = {\partial \over
\partial t}$) to emphasize the particular choice of variable. Given a
pseudodifferential operator
$$ A = \sum_{k= -\infty}^n a_k \partial^k \eqno(2.3)$$
we define, respectively, its differential and integral parts as
$$  A_+ = \sum^n_{k=0} a_k \partial^k, \qquad A_- =\sum^{-1}_{-\infty} a_k
\partial^k \eqno(2.4)$$
The residue and the trace of $A$ are, respectively,
$$ Res(A) = a_{-1} \eqno(2.5)$$
and
$$ Tr(A) = \int Res(A) dx = \int a_{-1} dx \eqno(2.6)$$

For the pseudodifferential operator given by (1.2), the hamiltonian flow,
defined by the second Gelfand-Dickey bracket,  generated by the functional
$$ F[u_i] = \int \epsilon_i(x) u_i(x) dx \eqno(2.7)$$
has the following operator form[3]
$$ \delta^{GD}_{\epsilon_i} L_n = (L_n V)_+ L_n -
L_n (V L_n)_+ \eqno(2.8)$$
where
$$ V = \partial^{-n+i-1} \epsilon_i(x) + \partial^{-n} q(x) \eqno(2.9)$$
with $q(x)$ satisfying
$$ Res[L_n, V] = 0 \eqno(2.10)$$
(2.10) is the consistency condition arising from the constraint $u_1=0$. For
reasons that will become clear later when i=2 (2.8) is called Virasoro
hamiltonian flow.

We shall study how the coefficient functions $u_i$'s transform under
diffeomorphism $x \rightarrow t$. The following terminologies are essential.
A function $f$ is called a primary of spin $h$ if, under $x \rightarrow t$, it
transforms as
$$ f(t) = \left({dx \over dt} \right)^h f(t) \eqno(2.11)$$
We denote by $F_h$ the space of all primaries of spin $h$. A pseudodifferential
operator $\Delta$ is called a covariant operator if it maps from $F_h$ to
$F_l$ for some $h$ and $l$. Symbollically, we denote
$$ \Delta \quad : \quad F_h \longrightarrow F_l \eqno(2.12)$$
We shall see later that  when $\Delta=L_n$ its form fixes $h$ and $l$
simultaneously. It is not hard to see that (2.12) is equivalent to
$$ \Delta(t) = \phi^{-l} \Delta(x) \phi^h \eqno(2.13)$$
where $\phi(x) = {dt \over dx}$.

We end this section by listing several elementary but useful properties of
primaries and covariant operators.

\noindent{(i)} If $f \in F_h$ and $g \in F_l$, then $fg \in F_{h+l}$.

\noindent{(ii)} If $\Delta_1 : F_h \rightarrow F_k$ and $\Delta_2 : F_k
\rightarrow F_l$, then $\Delta_2 \Delta_1 : F_h \rightarrow F_l$.

\noindent{(iii)} If $\Delta : F_h \rightarrow F_l$ and its inverse
$\Delta^{-1}$ exists, then $\Delta^{-1} : F_l \rightarrow F_h$

The proof of (iii) is simple. We just note that inverting (2.13) gives
$$ \Delta^{-1}(t)= \phi^{-h} \Delta^{-1}(x) \phi^{l} \eqno(2.14)$$
\medskip

\noindent{\bf III. \underbar{Diff($S^1$) Covariance and Virasoro Flow}}

Now we assume that the pseudodifferential operator $L_n$ can be covariantized,
i.e.
$$ L_n \quad : \quad F_h \longrightarrow F_l \eqno(3.1)$$
or, equivalently,
$$ L_n(t) = \phi^{-l} L_n(x) \phi^h \eqno(3.2)$$
for some $h$ and $l$. Quite obviously, not all values of $h$ and $l$ are
possible. To determine these values, let us recall that $u_1(t)$ and $u_1(x)$
must both vanish. However, the relation
$$ \partial_t = \phi^{-1}(x) \partial_x \eqno(3.3)$$
suggests that the leading term $\partial_t^n$ could possibly contribute to
the coefficient function $u_1(x)$. Indeed, for $n > 0$
$$\eqalign{\partial_t^n \phi^{-h} &= \phi^{-1} \partial_x \dots \phi^{-1}
\partial_x \phi^{-h}\cr
  &= \phi^{-(h+n)} \partial_x^n - [h+(h+1)+ \dots +(h+n-1)] \phi^{-(h+n+2)}
     \phi' \partial_x^{n-1} + \dots \cr}\eqno(3.4)$$
Hence, the covariance condition (3.2) requires
$$ h+(h+1)+ \dots + (h+n-1) = nh + {n(n+1) \over 2} = 0 \eqno(3.5)$$
and
$$ l= h+n \eqno(3.6)$$
It follows[17]
$$ h= - { n-1 \over 2}, \quad  l= {n+1 \over 2} \eqno(3.7)$$
For $n< 0$, the uses of
$$ \partial_t^{-1} = \partial_x^{-1} \phi \eqno(3.8)$$
also lead to the same conclusion. In other words, (3.7) is valid for all
nonzero $n$.

After imposing (3.7) the covariance  condition (3.2) reads
$$ L_n(t) \phi^{n-1 \over 2} = \phi^{-{n+1 \over 2}} L_n(x) \eqno(3.9)$$
which determines unambiguously how the coefficient functions $u_i$'s transform
under diffeomorphisms. The transformation of $u_2$ is of particular importance
as we shall see. Expanding the first two terms on the left hand side of (3.9)
gives
$$\partial_t^n \phi^{n-1 \over 2} = \phi^{-{n+1 \over 2}}
\big[\partial_x^n - { n(n^2-1) \over 12} \phi^2 \{\{x,t\}\} \partial_x^{n-2}
\big] + \dots \eqno(3.10)$$
and
$$ u_2(t) \partial_t^{n-2} \phi^{n-1 \over 2} = \phi^{-{n-3 \over 2}} u_2(x)
\partial_x^{n-2} + \dots \eqno(3.11)$$
where the schwartzian derivative $\{\{x,t\}\}$	is defined as
$$ \{\{x,t\}\} = {{{d^3 x} \over dt^3} \over {{dx} \over dt}} -
{3 \over 2} \left({{{d^2 x} \over dt^2} \over {{dx} \over dt}} \right)^2
\eqno(3.12)$$
Equating both sides of (3.9) now yields
$$ u_2(t) = u_2(x) \left({{dx} \over dt} \right)^2 + c_n \{\{x,t\}\}
\eqno(3.13)$$
where
$$ c_n = { n(n^2-1) \over 12} \eqno(3.14)$$
We therefore see that $u_2$ does not transform as a primary of spin 2 but has
an ``anomalous'' term proportional to the schwartzian derivative.
Since $u_2$ has
the same transformation law under diffeomorphisms as the energy-momentum tensor
in the conformal field theory, we call it the Virasoro generator.
We shall see in the next section that it  is the
presence of the anomalous term that enables us to decompose the functions
$u_i$'s into primaries.

The transformation laws for other coefficient functions can be worked out in
the same spirit. But the calculations are too tedious to be carried out here.
However, when we consider the infinitesimal form of diffeomorphisms the
corresponding transformation laws are quite manageable. Let
$$ t= x - \epsilon(x) \eqno(3.15)$$
then within linear approximation
$$ \phi(x) = {{dt} \over dx} = 1- \epsilon'(x) \eqno(3.16)$$
{}From (3.16) it is quite easy to show that (3.3) and (3.8) are equivalent to
$$\partial_t^{\pm 1} = \partial_x^{\pm 1} + [\partial_x^{\pm 1}, \epsilon]
\partial_x \eqno(3.17)$$
As a matter of fact, we can show easily by induction that (3.17) leads to
$$ \partial_t^i =\partial_x^i + [\partial_x^i, \epsilon] \partial_x
 \eqno(3.18)$$
for any integer $i$.
Since
$$\eqalign{ u_i(t) &= u_i(x-\epsilon(x)) + \delta_{\epsilon} u_i(x) \cr
  &= u_i(x) - \epsilon(x) u'_i(x) + \delta_{\epsilon} u_i(x), \cr}\eqno(3.19)$$
we have
$$ L_n(t) = L_n(x) - \epsilon(x)[ \partial_x, L_n(x) ] + [ L_n(x), \epsilon(x)]
\partial_x + \delta_{\epsilon} L_n(x)	\eqno(3.20)$$
On the other hand,
$$\big(1-\epsilon'(x) \big)^{-{{n+1} \over 2}} L_n(x) \big(1-\epsilon'(x)
\big)^
{-{{n-1} \over 2}} = L_n(x) + {{n+1} \over 2} \epsilon'(x) L_n(x) +
{{n-1} \over2}
 L_n(x)
\epsilon'(x)  \eqno(3.21)$$
Hence, (3.9), (3.20) and (3.21)  together imply
$$ \delta_{\epsilon} L_n(x) = {{n+1} \over 2} \epsilon'(x) L_n(x) + {{n-1}
 \over 2}
L_n(x) \epsilon'(x) + \epsilon(x) [\partial_x, L_n(x)] - [L_n(x), \epsilon(x)]
\partial_x \eqno(3.22)$$
(3.22) is a consequence of the covariance condition, which summaries the
transformation laws of $u_i$'s in the infinitesimal form.

Now we like to show that the transformation law (3.22) is nothing but the
hamiltonian flow (2.8) generated by the functional
$$ F_2[u_2] = \int \epsilon(x) u_2(x) dx \eqno(3.23)$$
First, we solve (2.10) for the above functional. The result is
$$ V = \partial_x^{-n} ({{n-1} \over 2}) \epsilon'(x) + \partial_x^{-n+1}
\epsilon(x) \eqno(3.24)$$
Simple algebras then give
$$\eqalign{ (L_n V)_+ &= {{n+1} \over 2} \epsilon'(x) + \epsilon(x)
\partial_x  \cr
   (V L_n)_+ &= -{{n-1} \over 2} \epsilon'(x) + \epsilon(x) \partial_x
\cr} \eqno(3.25)$$
Substituting (3.25) into (2.8) we obtain
$$\eqalign{\delta^{GD}_{\epsilon} L_n &= [{{n+1} \over 2}
\epsilon'(x) + \epsilon(x)
\partial_x] L_n(x) - L_n(x) [-{{n-1} \over 2} \epsilon'(x) +
\epsilon(x) \partial_x] \cr
&= {{n+1} \over 2} \epsilon'(x) L_n(x) + {{n-1} \over 2} L_n(x) \epsilon'(x) +
\epsilon(x) \partial_x L_n(x) - L_n(x) \epsilon(x) \partial_x \cr
&= {{n+1} \over 2} \epsilon'(x) L_n(x) + {{n-1} \over 2} L_n(x) \epsilon'(x)
+ \epsilon(x) [\partial_x, L_n(x) ] + [\epsilon(x), L_n(x) ] \partial_x \cr}
\eqno(3.26)$$
Note that (3.26) completely agrees with (3.22). We therefore have proved that
the infinitesimal form of the diff($S^1$) covariance of $L_n$ is equivalent to
the hamiltonian flow, defined by the second Gelfand-Dickey bracket, generated
by the functional given by (3.23).

{}From (3.26) we deduce with the help of (3.13)
$$\eqalign{\delta^{GD}_{\epsilon} u_2(x) &= \int \{u_2(x),u_2(y) \}_{GD}
\epsilon(y) dy \cr
&= u'_2(x) \epsilon(x) + 2 u_2(x) \epsilon'(x) + c_n \epsilon'''(x) \cr}
\eqno(3.27)$$
or, equivalently,
$$ \{ u_2(x), u_2(y) \}_{GD} = [u'(x) + 2u_2(x) \partial_x + c_n \partial_x^3]
\delta(x-y) \eqno(3.28)$$
(3.28) is the well-known Virasoro algebra. Because  of (3.28) we have called
the hamiltonian flow generated by the functional (3.23) the Virasoro
hamiltonian flow or simply Virasoro flow.

Before ending this section we simply remark that the above proof of equivalence
 applies without any change to the case when $L_n$ is just a differential
operator.
\medskip

\noindent{\bf IV. \underbar{Covariantization of $L_n$}}

In this section we shall covariantize the pseudodifferential operator $L_n$.
The key ingredient is the concept of covariant derivative. Given a
diffeomorphism $x \rightarrow v(x)$, we define
$$ b(x) = {{{d^2v} \over dx^2} \over {{dv} \over dx}}  \eqno(4.1)$$
It is a simple matter to check that under the diffeormorphism $x \rightarrow t$
this function transforms as
$$ b(t) = b(x) ({{dx} \over dt}) + {{d^2x} \over dt^2}
({{dx} \over dt})^{-1} \eqno(4.2)$$
One recognizes immediately that (4.2) is the transformation law for an
anomalous spin-1 primary. With the function $b$ we can define the covariant
derivative as
$$ D_k = \partial_x - k b(x)  \eqno(4.3)$$
For later uses, we also define
$$ D_k^l = D_{k+l-1} D_{k+l-2} \dots D_k  \qquad (l \geq 0) \eqno(4.4)$$
Using (4.2) and (3.3) we can easily show that under diffeomorphism
$$ D_k^l(t) = \phi^{-k-l} D_k^l(x) \phi^k  \eqno(4.5)$$
where, as usual $\phi = {{dt} \over dx}$. (4.5) means that the operator $D_k^l$
maps from $F_k$ to $F_{k+l}$. In other words, (4.3) and (4.4) really define
a series of covariant operators in the sense of  the definitions in Sec. II.
These covariant derivatives have been used in ref.[17] to covariantize the
differential operators. Since we are dealing with pseudodifferential operators,
we need something more. Obviously, what we need are the inverses of these
covariant operators, which have the following expressions
$$\eqalign{ D_k^{-1} &\equiv (\partial_x - k b)^{-1} = \partial_x^{-1} + k
\partial_x^{-1} b \partial_x^{-1} + k^2 \partial_x^{-1} b \partial_x^{-1} b
\partial_x^{-1} + \dots \cr
D_{k+l}^{-l} &\equiv (D_k^l)^{-1} = D_{k+1}^{-1} D_{k+2}^{-1} \dots
D_{k+l}^{-1}
\quad  (l \geq 1)} \eqno(4.6)$$
One should note that the subscript always denotes the spin of the domain space.
{}From Property (iii) in Sec. II it follows that these inverses are again
covariant derivatives.

For a given $L_n$ we shall choose $v(x)$ such that  in this particular
coordinate[17]
$$ u_2(v) = 0 \eqno(4.7)$$
By (3.13) it is equivalent to choosing $b$ to satisfy
$$ b'(x) - {1 \over 2} b(x)^2 = \{\{v,x\}\} = {{u_2(x)} \over c_n }
\eqno(4.8)$$
Clearly, such a choice is possible only when $c_n$ does not vanish. Hence, we
shall restrict ourselves to the cases $n \neq 0, \pm 1$.
Even when such a $b$ exists
it is not unique. We can replace $b$ by $b+\delta b$ as long as the schwartzian
 derivative $\{\{v,x\}\}$ is kept fixed. It amounts to requiring
$$ \delta b' =  b \delta b,$$
which has two more useful equivalent forms
$$ [ \partial_x -(k+1) b ] \delta b = \delta b (\partial_x - kb) \eqno(4.9)$$
and
$$ [ \partial_x -(k+1) b ]^{-1} \delta b = \delta b (\partial_x - kb)^{-1}
\eqno(4.10)$$
When $\delta b$ is constrained by (4.9) or (4.10) we can easily derive the
following useful formula
$$ \delta_b D_k^l = -{{l(l+2k-1)} \over 2} \delta b D_k^{l-1}
\qquad l=0, \pm 1,
\pm 2, \dots   \eqno(4.11)$$
{}From the covariance condition (3.9) it can be easily seen that the
differential
part and the integral part of $L_n$ transform independently; i.e.
$$ (L_n)_{\pm} (t) = \phi^{-{{n+1} \over 2}} (L_n)_{\pm} (x)
\phi^{-{{n-1} \over 2}}
\eqno(4.12)$$
Thus, both parts should be covariantized seperately. The covariantization of
$(L_n)_+ (n \geq 2)$ has been done in ref. [17] with the result
$$ (L_n)_+ = \Delta_2^{(n)} (u_2) + \sum^n_{k=3} \Delta_k^{(n)}(w_k,u_2)
\eqno(4.13)$$
where $w_k$ is a primary of spin $k$,
$$ \Delta_2^{(n)} (u_2) = D^n_{-{{n-1} \over 2}} =
(\partial_x - {{n-1} \over 2} b)
(\partial_x - {{n-3} \over 2} b) \dots
(\partial_x + {{n-1} \over 2} b) \eqno(4.14)$$

$$ \Delta_k^{(n)} (w_k, u_2) = \sum^{n-k}_{l=0} a_{k,l}^{(n)} \big[ D_k^l w_k
\big] D^{n-k-l}_{-{{n-1} \over 2}} \eqno(4.15)$$
and
$$ a_{k,l}^{(n)} = {{\pmatrix{ k+l-1 \cr l} \pmatrix{n-k \cr l}} \over
{\pmatrix{ 2k+l-1 \cr l}}}
\eqno(4.16)$$
The coefficients given by (4.16) are determined by requiring $\Delta_k^{(n)}$'s
 to depend on $b$ through the schwartzian derivative; i.e.
$$ \delta_b \Delta_k^{(n)} = 0 \eqno(4.17)$$
when $\delta b$ is subjected to (4.9).
To covariantize $(L_n)_-$, we naturally consider the covariant
pseudodifferntial operators
$$ \Delta_{n+k}^{(n)} (w_k,u_2) = \sum^{\infty}_{l=0} a^{(n)}_{n+k,l}
\big[D^l_{n+k} w_{n+k} \big] D^{-k-l}_{-{{n-1} \over 2}}  \quad k \geq 1
\eqno(4.18)$$
where $a^{(n)}_{n+k, 0} = 1$.
Requiring the dependence on $b$ only through the schwartzian derivative now
leads to the recursion relation
$$ a^{(n)}_{n+k,l} = - {{(k+l-1)(n+k+l-1)} \over {l(2n+2k+l-1)}}
a^{(n)}_{n+k,l-1} \eqno(4.19)$$
The solution to (4.19) is
$$ a^{(n)}_{n+k,l} =(-1)^l {{\pmatrix{k+l-1 \cr l} \pmatrix{n+k+l-1 \cr l}}
\over
{\pmatrix{2n+2k+l-1 \cr l}}} \eqno(4.20)$$
We have therefore obtained the desired covariant form
$$ (L_n)_- = \sum^{\infty}_{k=1} \sum^{\infty}_{l=0} a^{(n)}_{n+k,l}
\big[ D^l_{n+k} w_{n+k} \big] D^{-k-l}_{-{{n-1} \over 2}}  \eqno(4.21)$$
Working out explicitly the right hand side of (4.21) will yield the
decompositions of the form
$$ u_{n+k} = w_{n+k} + G_{n+k} (w_{n+k-1}, \dots, w_{n+1}, u_2) \eqno(4.22)$$
where $G_{n+k}$ is a differential polynomial in
$u_2,w_{n+1}, \dots, w_{n+k-1}$ and their derivatives.
Inverting (4.22) gives the definitions of primaries in terms of
coefficient functions
$$ w_{n+k} = u_{n+k} + H_{n+k} (u_{n+k-1}, \dots, u_{n+1}, u_2) \eqno(4.23)$$
where $H_{n+k}$ is again a differential polynomial. We thus conclude that the
primaries of spin higher than or equal to n+1 can be defined from the
coefficient functions. Combining this with the result of Sec. III
we deduce that
a primary basis for the algebra $W_{KP}^{(n)}$ ($n \geq 2$) indeed exists.

Next, we consider $L_{-n}$ with $n \geq 2$. Following the same steps we obtain
$$ L_{-n} = \Delta_2^{(-n)} (u_2) + \sum^{\infty}_{k=3} \Delta_k^{(-n)} (w_k,
u_2),  \eqno(4.24)$$
where
$$\eqalignno{\Delta_2^{(-n)}(u_2) &= (\partial_v^n)^{-1} =
D^{-n}_{{{n+1} \over 2}}
 \qquad &(4.25) \cr
\Delta_k^{(-n)}(w_k, u_2) &= \sum^{\infty}_{k=3} a^{(-n)}_{k,l} \big[D^i_k w_k
\big] D^{-n-k-l}_{{{n+1} \over 2}} \qquad &(4.26) \cr}$$
and
$$ a^{(-n)}_{k,l} = (-1)^l {{\pmatrix{k+n+l-1 \cr l} \pmatrix{k+l-1 \cr l}}
\over
{\pmatrix{2k+l-1 \cr l}}} \quad (k \geq 3) \eqno(4.27)$$
The conclusion is the same. For $n \geq 2$, $L_{-n}$ can be covariantized and
therefore the algebra $W_{KP}^{(-n)}$ has a primary basis.

Several remarks are  in order. First, one sees that (4.16), (4.20) and (4.27)
can be summaried by a single formula
$$ a^{(n)}_{k,l} = (-1)^l {{\pmatrix{k-n+l+1 \cr l} \pmatrix{k+l-1 \cr l}}
\over
{\pmatrix{2k+l-1 \cr l}}} \quad (k \geq 3, l \geq 0) \eqno(4.28)$$
Note, in particular, that $a_{k,l}^{(n)}$ given by (4.28) vanishes when
$n \geq k \geq 3 $
and $l > n-k$. This is consistent with the fact that the right hand side of
(4.15) is just a finite summation.
Secondly, using
$$ D^k_{p+q} = \sum^{\infty}_{l=0} \pmatrix{k \cr l} [D^l_p w_p] D^{k-l}_q
\eqno(4.29)$$
which is the covariant analogue of (2.1), we observe that for $\mid n \mid
\geq 2$
$$ \Delta_2^{(n)}(u_2) \Delta_k^{(-n)} (w_k,-u_2) \Delta_2^{(n)} (u_2) =
 \sum^{\infty}_{l=0} b_{k,l}^{(n)} [D_k^l w_k ] D^{n-k-l}_{-{{n-1} \over 2}}
\eqno(4.30)$$
where
$$ b^{(n)}_{k,l} = \sum^{\infty}_{p=0} \pmatrix{n \cr p} a_{k,l-p}^{(n)} \qquad
\big( a^{(n)}_{k,l} = 0  \quad if \quad l<0 \big) \eqno(4.31)$$
One should notice that the sign of $u_2$ in the expression of $\Delta_k^{(-n)}$
has been reversed in order to assure that the same central charge $c_n$ is used
in (4.8). Since $b^{(n)}_k = a_k^{(n)} =1$ and since the right hand side of
(4.30) must depend on $b$ only through the schwartzian derivative as the left
hand side does, we deduce
$$ a^{(n)}_{k,l} = b^{(n)}_{k,l} = \sum^{\infty}_{p=0} \pmatrix{n \cr p}
a^{(-n)}_{k,l-p} \eqno(4.32)$$
and
$$\Delta_k^{(n)} (w_k, u_2) = \Delta_2^{(n)} (u_2) \Delta_k^{(-n)} (w_k, -u_2)
\Delta_2^{(n)} (u_2) \eqno(4.33)$$
(4.32) is not a obvious relation. We did explicitly verify it for a few simple
cases even though we have not devised a direct proof.  Finally, we like to
remark that the covariant operator $\Delta^{(n)}_k$ has been chosen to depend
linearly on $w_k$ and its derivative. In general, we can add terms which are
multilinear in $w_j$'s and their derivatives. Quite clearly, adding
$$ \sum^{\infty}_{l_{\scriptstyle 1}, \dots, l_{\scriptstyle p} =0}
a^{(n)}_{k_{\scriptstyle 1}, \dots ,k_{\scriptstyle p} ; l_{\scriptstyle 1},
\dots, l_{\scriptstyle p}} [D_{k_1}^{l_1} w_{k_1}] \dots [D^{l_p}_{k_p}
w_{k_p}] D^{n-k_1- \dots -k_p -l_1 \dots \l_P}_{-{{n-1} \over 2}}
\eqno(4.34)$$
where $k_i \geq 3$ and $k_1 + \dots +k_p = k$, to $\Delta^{(n)}_k$ and choosing
the coefficients properly  would just change the form of the differential
polynomials in (4.22) and (4.23). However, as one can see easily that the
choice	of the set of coefficients is not unique now, we will not discuss any
further. Nevertheless, we will come back to this point when we try to
explicitly decompose $u_k$'s into a differential  polynomial in primaries and
their derivatives in the next section.
\medskip

\noindent{\bf V. \underbar{Primaries in $W^{(n)}_{KP}$}}

In this section we shall use the Virasoro hamiltonian flow (3.26) to decompose
explicitly, for $n \ge 2$, the coefficient functions $u_{n+1}, \dots , u_{n+4}$
into differential polynomials in $w_{n+1}, \dots, w_{n+4}$ and $u_2$. There are
two purposes for these calculations. First, we recall that when $n > 0$  the
coefficient functions $\{ u_j \}_{j \ge n+1}$ generate a diff($S^1$) submodule.
 However, (4.23) shows that in defining the primaries $w_{n+k}$'s we must
introduce $u_2$ which belongs to the diff($S^1$) submodule generated by
$\{u_2,
\dots, u_n \}$. Hence , it is worth while to do the docompostions directly from
the Virasoro hamiltonian flow in order to see why $u_2$ must appear. Secondly,
these explicit expressions will be compared with those following from (4.21)
to serve as an independent verification of our results.

With some straightforward algebras we obtain from (3.26)
$$ \delta_{\epsilon} u_{n+p} = \epsilon u'_{n+p} + (n+p) \epsilon' u_{n+p}
 +\sum^{p-1}_{k-1} \bigl[{{n-1} \over 2}
\pmatrix{-k \cr p-k} - \pmatrix{-k \cr p-k+1} \bigr]
\epsilon^{(p-k+1)} u_{n+k}  \eqno(5.1)$$
The first four transformation laws from (5.1) are
$$\eqalignno{\delta_{\epsilon} u_{n+1} &= \epsilon u'_{n+1} +
(n+1) \epsilon' u_{n+1}
 &(5.2)\cr
\delta_{\epsilon} u_{n+2} &= \epsilon u'_{n+2} + (n+2) \epsilon' u_{n+2}
-{{n+1} \over 2} \epsilon'' u_{n+1}  &(5.3)\cr
\delta_{\epsilon} u_{n+3} &= \epsilon u'_{n+3} + (n+3) \epsilon' u_{n+3}
- (n+2) \epsilon'' u_{n+2} + {{n+1} \over 2} \epsilon''' u_{n+1}  &(5.4)\cr
\delta_{\epsilon} u_{n+4} &= \epsilon u'_{n+4} + (n+4) \epsilon' u_{n+4}
-{{3(n+3)} \over 2} \epsilon'' u_{n+3} + {{(3n+5)} \over 2} \epsilon''' u_{n+2}
\cr
{\rm }\qquad \qquad &- {n+1 \over 2} \epsilon'''' u_{n+1}  &(5.5)\cr}$$
It is an immediate consequence of (5.2) that $u_{n+1}$ is a primary of spin
$n+1$. We thus write
$$ u_{n+1} = w_{n+1} \eqno(5.6)$$
Next, we define
$$ w_{n+2} = u_{n+2} + \alpha_2 w'_{n+1}  \eqno(5.7)$$
Using (5.2), (5.3) and (5.6) we find
$$\delta_{\epsilon} w_{n+2} = \epsilon w'_{n+2} + (n+2) \epsilon' w_{n+2}
 + (\alpha_2 - {1 \over 2} ) (n+1) \epsilon'' w_{n+1} \eqno(5.8)$$
To make $w_{n+2}$ a primary we must set
$$ \alpha_2 = {1 \over 2}, \eqno(5.9)$$
which then gives the decomposition
$$ u_{n+2} = w_{n+2} - {1 \over 2} w'_{n+1} \eqno(5.10)$$
Now let
$$ w_{n+3} = u_{n+3} + \alpha_3 w'_{n+2} + \beta_3 w''_{n+1} \eqno(5.11)$$
The transformation law for $w_{n+3}$ reads
$$ \eqalign{\delta_{\epsilon} w_{n+3} &= \epsilon w'_{n+3}
+ (n+3) \epsilon' w_{n+3}
+ (\alpha_3 -1)(n+2) \epsilon'' w_{n+2} \cr
{\rm } \qquad &+ [\beta_3 (2n+3) +{1 \over 2} (n+2) ]
\epsilon'' w'_{n+1}
+ (\beta_3 + {1 \over 2} )(n+1) \epsilon''' w_{n+1}\cr}
\eqno(5.12)$$
Clearly, no choice of $\alpha_3$ and $\beta_3$ can simultaneously make the
last three terms on the right hand side of (5.12) vanish. In other words,
 the primary $w_{n+3}$ can not be defined as a differential polynomial	in
$u_{n+1}, u_{n+2}$ and $u_{n+3}$. However, we observe that $\epsilon'''$ is
proportional to the anomalous term in $\delta_{\epsilon} u_2$. This suggests
a way out. We choose $\alpha_3$ and $\beta_3$ to make the first two terms
vanish and then add a term $\gamma_3 (u_2 w_{n+1})$ to the right hand side of
(5.11) to take care of the last term. Indeed, from the transformation law
$$\delta_{\epsilon} (u_2 w_{n+1}) = \epsilon (u_2 w_{n+1})' + (n+3) \epsilon'
 (u_2 w_{n+1}) + c_n \epsilon''' w_{n+1} \eqno(5.13)$$
we see that the addition of the new term would simply add $c_n \gamma_3
\epsilon''' w_{n+1}$ to the right hand side of (5.12). Hence, the choice
$$ \alpha_3 =1, \qquad \beta_3 =- { n+2 \over {2(2n+3)}},\qquad
 \gamma_3 = - { (n+1)^2 \over {2 c_n (2n+3)} } \eqno(5.14)$$
makes $w_{n+3}$
a real primary and thus the decomposition of $u_{n+3}$ is
$$ u_{n+3} = w_{n+3} - w'_{n+2} + { n+2 \over {2(2n+3)}} w''_{n+1} +
 {(n+1)^2 \over { 2 c_n (2n+3)}} u_2 w_{n+1} \eqno(5.15)$$
This analysis shows that when $n=1$ the primary $w_{n+3} = w_4$ can not be
defined due to the fact that $c_n = c_1 = 0$ (equivalently, $u_2$ does not
transform anomalously)[18].
It amounts to saying that the algebra $W_{KP}^{(1)}$,
which the second hamiltonian structure of the KP hierarchy, has no primary
basis.

Finally, we consider
$$\eqalign{ w_{n+4} &= u_{n+4} + \alpha_4 w'_{n+3} + \beta_4 w''_{n+2}
+ \gamma_4 w'''_{n+1} + \mu_4 u_2 w_{n+2} + \cr
 &\qquad \xi_4 u_2 w'_{n+1} +  \eta_4 u'_2 w'_{n+1} \cr}\eqno(5.16)$$
whose transformation law reads
$$\eqalign{\delta_{\epsilon} w_{n+4} &= \epsilon w'_{n+4} + (n+4) \epsilon'
w_{n+4} + \epsilon'' \big( A w_{n+3} + B w'_{n+2} + C w''_{n+1} + D u_2 w_{n+1}
\big) \cr
&\qquad \epsilon''' \big( E w_{n+2} + F w'_{n+1} \big) + G
\epsilon'''' w_{n+1} \cr}\eqno(5.17.1)$$ where
$$\eqalignno{ A &= (\alpha_4 - {3 \over 2})(n+3)  &(5.17.2)\cr
B &= (2n+5) \beta_4 + {{3(n+3)} \over 2}  &(5.17.3)\cr
C &= (3n+6) \gamma_4 - { 3(n+2)(n+3) \over {4(2n+3)} } &(5.17.4) \cr
D &= 2 \eta_4 + (n+1) \xi_4 - {3(n+3)(n+1)^2 \over {4 c_n (2n+3)}} &(5.17.5)
\cr
E &= (n+2) \beta_4 + c_n \mu_4 + {{(3n+5)} \over 2}  &(5.17.6)\cr
F &= (3n+4) \gamma_4 + c_n \xi_4 - {{(3n+5)} \over 4}	&(5.17.7)\cr
G &= (n+1) \gamma_4 + c_n \eta_4 - {{(n+1)} \over 2}   &(5.17.8)\cr}$$
Demanding $A=B=C=D=E=F=G=0$ gives seven equations for only six unknowns.
If we drop $D=0$ for a moment and solve the other six equations, we get
$$\eqalign{\alpha_4 &= {3 \over 2}, \quad \beta_4= -{3(n+3) \over {2(2n+5)}},
\quad \gamma_4 = { n+3 \over {4(2n+3)}} \cr
\mu_4 &= -{(n+1)(3n+7) \over{2 c_n (2n+5)}}, \cr
\eta_4 &= \xi_4 = {3(n+1)^2 \over {4 c_n (2n+3)}} \cr}\eqno(5.18)$$
Substituting (5.18) back into (5.17.5)
we find  that $D=0$ is indeed satisfied.
Therefore, with the coefficients given by (5.18), (5.16)  defines a
primary. The decomposition for $u_{n+4}$ is then
$$\eqalign{ u_{n+4} &= w_{n+4} - {3 \over 2} w'_{n+3} + {3(n+3) \over
{2(2n+5)}}
w''_{n+2} - {n+3 \over {4(2n+3)}} w'''_{n+1} \cr
&\qquad + {(n+1)(3n+7) \over{2 c_n (2n+5)}} u_2 w_{n+2} - {3(n+1)^2 \over {4
c_n (2n+3)}} (u_2 w_{n+1})' \cr}\eqno(5.19)$$
We should know that the decomposition (5.19)  is by no means unique. A
redefinition like
$$ w_{n+4} \longrightarrow  w_{n+4} + w_3 w_{n+1}   \eqno(5.20)$$
is certainly allowed. Redefinitions of this sort amount to introducing terms
bilinear in $w_i$'s and their derivatives in the decompositions. As we remarked
in Sec. IV., more generaly, terms which are multilinear in $w_i$'s and their
derivatives can be introduced. $n+4$ is the lowest spin where the arbitrariness
of this type shows up.

Now we like to compare the above explicit decomposition formulae with those
resulting form (4.21). The needed formulae are
$$\eqalign{D_k w_k &= w'_k - kb w_k \cr
D_k^2 w_k &= w''_k -(2k+1) b w'_k -kb' w_k +k(k+1) b^2 w_k \cr
D_k^3 w_k &= w'''_k -3k b w''_k -(3k+1) b' w'_k +(13k^2+6k+2) b^2 w'_k \cr
    &\quad -kb'' w_k + k(3k+4)bb' w_k -k(k+1)(k+2) b^3 w_k \cr}\eqno(5.21)$$
and
$$\eqalign{D_k^{-1} &= \partial_x^{-1} + (k-1)b \partial_x^{-2} -
 [(k-1)b' - (k-1)^2 b^2 ] \partial_x^{-3} \cr
&\qquad +[(k-1)b'' -3(k-1)^2bb' +(k-1)^3 b^3] \partial_x^{-4} + \dots \cr
D_k^{-2} &= \partial_x^{-2} + (2k-1) b \partial_x^{-3} + [-(3k-4)b' +
(3k^2-9k+7)b^2] \partial_x^{-4} + \dots \cr
D_k^{-3} &= \partial_x^{-3} + 3(k-2) b \partial_x^{-4} + \dots \cr
D_k^{-4} &= \partial_x^{-4} + \dots \cr}\eqno(5.22)$$
After some algebras we obtain
$$\eqalign{\Delta_{n+1}^{(n)} (w_{n+1},u_2) &= w_{n+1} \partial_x^{-1} -
{1 \over 2} w'_{n+1} \partial_x^{-2} + \big[{n+2 \over{2(2n+3)}} w''_{n+1} +
 {(n+1)^2 \over {2 c_n (2n+3)}} u_2 w_{n+1} \big] \partial_x^{-3} \cr
&\qquad + \big[ -{n+3 \over {4(2n+3)}} w'''_{n+1} - {3(n+1)^2 \over {4c_n
(2n+3)}} (u_2 w_{n+1})' \big] \partial_x^{-4} + \dots \cr
\Delta_{n+2}^{(n)} (w_{n+2},u_2) &= w_{n+2} \partial_x^{-2} + \big[
{3(n+3) \over {2(2n+5)}} w''_{n+2} + {(n+1)(3n+7) \over {2 c_n (2n+5)}} u_2
w_{n+2} \big] \partial_x^{-4} + \dots \cr
\Delta_{n+3}^{(n)} (w_{n+3},u_2) &= w_{n+3} \partial_x^{-3} -{3 \over 2}
w'_{n+2} \partial_x^{-4} + \dots \cr
\Delta_{n+4}^{(n)} (w_{n+4},u_2) &= w_{n+4} \partial_x^{-4} + \dots \cr}
\eqno(5.23)$$
It is a simple matter to check that (5.23) completely agrees with (5.6),
(5.10), (5.15) and (5.19). This completes our comparison.
\medskip

\noindent{\bf VI. \underbar{Discussions and Conclusions}}

We have shown that when $n \ne 0, \pm 1$ the pseudodifferential operators
given  by (1.2) can be covariantized,  that  is, the coefficient functions
can be decomposed into differential polynomials in primaries and thier
derivatives. This therefore gives a proof for the existence of primary basis
for the corresponding algebra $W_{KP}^{(n)}$. For $n \geq 2$ we have worked
out the decompositions of $\{u_j\}_{ n+4 \geq j \geq n+1}$ from their
transformationn laws under Virasoro hamiltonian flow. The results completely
agree with those from diff($S^1$) covariantization of $L_n$.

Two possible generizations of the constructions in this paper are worth
mentioning. First, we may consider the	pseudodifferential operator of a
complex degree $q$.  For such a case  it is necessary to define an object like
$D_k^{q}$, which is a covariant operator with a complex power. Once this
is done, an analogous  construction then can be carried out. Secondly, we may
consider the super-pseudodifferential operators in which the derivative
$\partial_x$ is replaced by the super-derivative $D=\partial_{\theta} + \theta
 \partial_x$ in the $(1,1)$ superspace and the coefficient functions $\{u_i\}$
by the superfields $\{U_i = v_i + \theta w_i\}$. Works in these two directions
are now in progress.

\vskip .5cm
\noindent{\bf \underbar{Acknowledgement}}

The author likes to thank Prof. H.C. Yen for encouragement. This work
was supported by the National Science Council of Republic of China
under Grant No. NSC-83-028-M-007-008.

\vfil\eject
\noindent{\bf \underbar{References}}

\vskip .5cm
\item{[1]} P.D. Lax, Comm Pure Appl. Math. {\bf 21}, 467 (1968).
\item{[2]} I.M. Gelfand and L.A. Dickey, Func. Anal. Appl. {\bf 10},
	   4 (1976).
\item{[3]} M. Adler, Invent. Math. {\bf 50}, 219 (1979).
\item{[4]} A.B. Zamolodchikov, Theor. Math. Phys. {\bf 65}, 1205 (1985).
\item{[5]} I. Bakas, Nucl. Phys. {\bf B302}, 189 (1988); Phys. Lett.
{\bf B213}, 313 (1988).
\item{[6]} Y. Watanabe, Lett. Math. Phys. {\bf 7}, 99 (1983).
\item{[7]} L.A. Dickey, Ann. NY Acad. Sci. {\bf 491}, 131 (1987).
\item{[8]} K. Yamagishi, Phys. Lett. {\bf B259}, 436 (1991).
\item{[9]} F. Yu and Y.-S. Wu, Phys. Lett. {\bf B236}, 220 (1991).
\item{[10]} J.M. Figueroa-O'Farrill, J. Mas and E. Ramos, Phys. Lett.
{\bf B266}, 298 (1991).
\item{[11]} A. Das, W.-J. Huang and S. Panda, Phys. Lett. {\bf B271},
109 (1991).
\item{[12]} A.O. Radul, in: Applied methods of nonlinear analysis and
control, eds. A Mironov, V. Moroz and M. Tshernjatin (MGU, Moscow
1987) [in Russian].
\item{[13]} A. Das and W.-J. Huang, J. Math. Phys. {\bf 33}, 2487 (1992).
\item{[14]} J.M. Figueroa-O'Farrill, J. Mas and E. Ramos, A one-
parameter family of hamiltonian structures for the KP hierarchy and a
continuous deformation of the $W_{KP}$ algebra, preprint BONN-HE-92-20,
US-FT-92/7, KUL-TF-92/20.
\item{[15]} P. Mathieu, Phys. Lett. {\bf B208}, 101 (1988).
\item{[16]} Q. Wang, P.K. Panigraphi, U. Sukhatme and W.-K. Keung,
Nucl. Phys. {\bf B344}, 194 (1990).
\item{[17]} P. Di Francesco, C. Itzykson, and J.-B. Zuber, Comm. Math.
Phys. {\bf 140}, 543 (1991).
\item{[18]} J.M. Figueroa-O'Farrill, J. Mas and E. Ramos, Phys. Lett.
{\bf B299}, 41 (1993).
\item{[19]} J.M. Figueroa-O'Farrill and E. Ramos, Phys. Lett. {\bf B262},
265 (1991).

\bye

 \end